\documentclass[12pt,preprint]{aastex}

\usepackage{epsfig}

\begin{document}
%\twocolumn[

\title{Nonthermal radiation from clusters of galaxies:\\ 
the role of merger shocks in particle acceleration}

\author{Stefano Gabici}
\affil{Dipartimento di Astronomia e Scienza dello Spazio, Universit\`a 
di Firenze,\\Largo E. Fermi 5, I-50125 Firenze, Italy} 
\email{gabici@arcetri.astro.it}
\author{Pasquale Blasi}
\affil{INAF/Osservatorio Astrofisico di Arcetri,\\
Largo E. Fermi 5, I-50125 Firenze, ITALY}
\email{blasi@arcetri.astro.it}
\begin{abstract}
Nonthermal radiation is observed from clusters of galaxies in the radio,
hard X-rays, and possibly in the soft X-ray/UV bands. While it is known
that radiative processes related to nonthermal electrons are responsible
for this radiation, the sites and nature of particle acceleration are not
known. We investigate here the acceleration of protons and electrons in the 
shocks originated during mergers of clusters of galaxies, where the Fermi
acceleration may work. We propose a semi-analytical model to evaluate the 
Mach number of the shocks generated during clusters mergers and we use this
procedure to determine the spectrum of the accelerated particles
for each one of the shocks produced during the merger history of a cluster.
We follow the proton component, accumulated over cosmological time scales,
and the short lived electron component. We conclude that efficient particle 
acceleration, resulting in nonthermal spectra that compare to observations,
occurs mainly in minor mergers, namely mergers between clusters with very 
different masses. Major mergers, often invoked to be sites for the production of 
extended radio halos, are found to have on average too weak shocks and are
unlikely to result in appreciable nonthermal activity.
\end{abstract}
%]
\section{Introduction}\label{sec:intro}
Rich clusters of galaxies are strong X-ray sources with luminosity typically 
in the range $L_X\sim 10^{43}-10^{45} erg/s$. The X-ray emission is
well explained as bremsstrahlung radiation of the very hot ($T \sim 10^8 K$), 
low density ($n_e \sim 10^{-3} cm^{-3}$) and highly ionized intracluster 
electron gas. 

There is now compelling evidence for the existence, besides the
thermal electron gas, of a nonthermal population of particles, responsible for
extended synchrotron radio halos in a growing fraction of the observed 
clusters (see Feretti et al., 2000 for a recent rewiew), as well as for
hard X-ray (HXR) and extreme ultra violet (EUV) excesses detected in a few 
clusters (see e.g. Fusco Femiano et al., 1999, 2000; Lieu et al., 1996).
While it is clear that radio emission is due to synchrotron radiation from 
relativistic electrons, it is not as clear how these particles are accelerated.
Several models have been proposed, based on shock acceleration of electrons 
in merger shocks (Roettiger, Burns \& Stone, 1999; Sarazin 1999; Takizawa \&
Naito, 2000; Fujita \& Sarazin 2001) or models in which electrons are secondary 
products of hadronic interactions (Dennison 1980, Colafrancesco \& Blasi 1998;
Blasi \& Colafrancesco 1999; Dolag \& Ensslin 2000)
and finally models in which electrons are continuously reenergized by turbulence
(Schlickeiser, Sievers, \& Thiemann 1987; Brunetti et al. 2001; 
Ohno, Takizawa, \& Shibata 2002).
 
HXR and EUV radiation in excess of the thermal emission may be generated by
inverse Compton scattering (ICS) of relativistic electrons off the photons of 
the cosmic microwave background radiation. When applied to the Coma cluster,
these models require values of the volume averaged magnetic field which are 
smaller than those measured through Faraday rotation, which are typically 
of several $\mu G$ (Eilek 1999; Clarke, Kronberg \& B\"{o}ringer 1999). 
This conclusion can be possibly avoided only
by constructing models in which a cutoff in the electron spectrum
is tuned up in order to reduce the corresponding synchrotron emission. In
these cases the magnetic field can be as high as $0.3-0.4\mu G$ (Brunetti et
al. 2001). In the case of a secondary origin for the radiating
electrons, the small magnetic fields imply a large cosmic ray content in the
intracluster gas. In the case of the Coma cluster, the gamma ray upper limit
found by Sreekumar et al. (1996) is exceeded by the gamma ray flux from the decay 
of neutral pions, as shown by Blasi \& Colafrancesco (1999).

The HXR excess might also be the result of bremsstrahlung emission from a 
population of thermal electrons whose distribution function is slightly 
different from a Maxwell-Boltzmann (MB) distribution (Ensslin, Lieu \&
Biermann 1999; Blasi 2000; Dogiel 2000; Sarazin \& Kempner 2000). A tail 
might in fact be induced in the MB distribution by the presence of MHD waves 
that resonate with part of the thermal electrons (Blasi, 2000). This model 
requires an energy input comparable with the energy budget of a cluster 
merger, and implies a substantial heating of the intracluster gas (this was 
shown by Blasi (2000) by solving the full Fokker-Planck equations, including
Coulomb scattering). If the process lasts for too long a time (larger than a 
few hundred million years) the cluster is heated to a temperature well in 
excess of the observed ones, and the model fails. In this case the arguments 
presented by Petrosian (2001) apply. 

The presence of tails in the MB electron distribution can be tested through
observations of the Sunyaev-Zeldovich (SZ) effect, as proposed by Blasi, 
Olinto \& Stebbins (2000) (see Ensslin \& Kaiser (2000) for a general 
discussion of the SZ effect including nonthermal effects).
Clearly, by simply observing radio radiation and hard X-ray radiation from
clusters, it is extremely difficult, if not impossible to discriminate among
classes of models. The study of the SZ effect allows one to partly break the 
degeneracy. An even more powerful tool is represented by gamma ray astronomy.
Some of the models in the literature predict gamma ray emission to some
extent, while others (this is the case of nonthermal tails in the MB 
distribution) do not make precise predictions about the gamma ray emission,
and in fact do not require it. Clusters of galaxies are among the targets for
observations by the GLAST satellite. These observations will open a new
window onto the nonthermal processes occurring in the intracluster gas,
and will allow one to understand the origin of the observed radiation at lower
frequency (for a recent review see (Blasi 2002)).

As stressed above, there is at present no compelling evidence in favor of
any of the proposed acceleration sites for the nonthermal particles in 
clusters. Nevertheless, energetic events in the history of a cluster
represent good candidates, and cluster mergers, that build up the cluster 
itself hierarchically, fit the description. It seems therefore reasonable
to associate the existence of nonthermal particles to some process occurring
during these mergers. This argument is made stronger by the fact that mergers
are also thought to be responsible for the heating of the intracluster gas. 
A possible observational evidence for a correlation between major mergers and
radio halos has recently been found by Buote (2001). In particular, the 
correlation exists between the radio emission at 1.4 GHz and the degree of 
departure from virialization in the shape of clusters, interpreted as a 
consequence of a recent or ongoing merger that visibly changed the dark matter 
distribution in the cluster core.

The shocks that are formed in the baryon components of the merging
clusters are able to convert part of the gravitational energy of the system
into thermal energy of the gas, as shown by direct observations (e.g. 
(Markevitch, Sarazin and Vikhlinin 1999)). It has been claimed that if these
shocks are strong enough, they can efficiently accelerate particles by first
order Fermi acceleration (Fujita \& Sarazin 2001; Miniati et. al. 2001a,b; 
Blasi 2000). 
The consequences of these shocks on the nonthermal content of
clusters of galaxies may be dramatic, and deserve to be considered in detail.
Both electrons and protons (or nuclei) are accelerated at the shock surfaces
during mergers, but the dynamics of these two components is extremely 
different: high energy electrons have a radiative lifetime much shorter than 
the age of the cluster, so that they rapidly radiate most of their energy away
and eventually pile up at lorentz factors $\sim 100$. On the other hand,
protons lose only a small fraction of their energy during the lifetime of the
cluster, and their diffusion time out of the cluster are even larger, so that 
they are stored in clusters for cosmological times (Berezinsky, Blasi \&
Ptuskin 1997; Volk, Aharonian, \& Breitschwerdt 1996). In other words, while
for high energy electrons only the recent merger events are important to 
generate nonthermal radiation that we can observe, in order to determine
the proton population of a cluster (that can generate secondary electrons)
we need to take into account the all history of the cluster. 

In this paper we simulate merger histories of galaxy clusters and we calculate
the properties of the shocks generated during the merger events and the related
spectrum of particles accelerated at the shocks. We also account for the 
re-energization of particles preexisting within the merging clusters. We 
demonstrate that major mergers (mergers between clusters with comparable
masses) which are supposed to be the most energetic events and that are often
thought to be responsible for nonthermal activity, generate shocks that are 
typically weak and cannot account for the spectral slopes of the observed 
nonthermal radiation. 
Shocks in small mergers are also considered, and we find that they
may play an important role to generate the nonthermal radiation by primary 
electrons accelerated recently. As far as the proton component is concerned,
the time integrated spectra are energetically dominated by major mergers, so
that the resulting spectra are very steep, due to the weakness of the 
corresponding shocks.

Throughout the paper we assume a flat cosmology ($\Omega_0=1$) 
with $\Omega_m=0.3$, $\Omega_{\Lambda}=1-\Omega_m=0.7$ and a value for the 
Hubble constant of $70\, \rm{km/s/Mpc}$.

The paper is organized as follows: in \S \ref{sec:simula} we discuss our 
simulations for the reconstruction of the merger tree; in \S \ref{sec:accelera} 
we discuss the basics of shock acceleration and reacceleration in clusters of 
galaxies, and the physics of cosmic ray confinement. In \S \ref{sec:results}
we describe our results, and we conclude in \S \ref{sec:conclude}.

\section{Constructing the merger tree of a cluster\label{sec:simula}}

The standard theory of structure formation predicts that larger structures 
are the result of the mergers of smaller structures: this hierarchical model
of structure formation in the universe has been tested in several independent
ways and provides a good description of the observations of the mass function 
of clusters of galaxies and their properties.

While a complete understanding of the process of structure formation can only 
be achieved by numerical N-body simulations, an efficient and analytical 
description can also be achieved. It represents an extremely
powerful tool in that it allows one to reconstruct realizations of the merger
history of a cluster with fixed mass at the present time. These analytical 
descriptions come in different flavors and are widely discussed in the 
literature. Historically, the first approach to the problem was proposed 
by Press \& Schechter (1974, hereafter PS) and successively developed by 
Bond et al. (1991) and Lacey \& Cole (1993, hereafter LC) among others. 
In the PS formalism, the differential comoving number density of clusters with 
mass $M$ at cosmic time $t$ can be written as:
\begin{equation}
\frac{dn(M,t)}{dM}=\sqrt{\frac{2}{\pi}}\,\frac{\varrho}{M^2}\,
\frac{\delta_c(t)}
{\sigma(M)}\,\left|{\frac{dln \sigma(M)}{dln M}}\right| exp\left[-\frac
{\delta_c^2(t)}{2\sigma^2(M)}\right].
\end{equation}
The rate at which clusters of mass $M$ merge at a given time $t$ is written 
as a function of $t$ and of the final mass $M^{\prime}$ (LC, 1993):

$$R(M,M^{\prime},t)dM^{\prime}=$$
\begin{eqnarray}
\sqrt{\frac{2}{\pi}}\,\left|
\frac{d\delta_c(t)}{dt}\right|\,\frac{1}{\sigma^2(M^{\prime})}\,
\left|\frac{d\sigma(M^{\prime})}{dM^{\prime}}\right|\,
\left(1-\frac{\sigma^2(M^{\prime})}{\sigma^2(M)}\right)^{-3/2} \nonumber \\
\rm{exp}\left[-\frac{\delta_c^2(t)}{2}\left(\frac{1}{\sigma^2(M^{\prime})}-
\frac{1}{\sigma^2(M)}\right)\right]dM^{\prime},
\end{eqnarray}
where $\varrho$ is the present mean density of the universe, $\delta_c(t)$ 
is the critical density contrast linearly extrapolated to the present time 
for a region that collapses at time $t$, and $\sigma(M)$ is the current rms 
density fluctuation smoothed over the mass scale $M$.
For $\sigma(M)$ we use an approximate formula proposed by Kitayama (1997), 
normalized by assuming a bias parameter $b=0.9$. We adopt the expression of
$\delta_c(t)$ given by Nakamura \& Suto (1997). In this respect our approach
is similar to that adopted by Fujita \& Sarazin (2001).

Salvador-Sol\'e, Solanes \& Manrique (1998) modified the model illustrated
above, by introducing a new parameter,  $\Delta_m = r_{crit}= 
[(M^{\prime}-M)/M]_{crit}$, defined as a peculiar value of the captured
mass that separates the accretion events from merger events. 
Events in which a cluster of mass $M$ captures a dark matter halo with 
mass smaller then $\Delta_m M$ are considered as continuous mass 
accretion, while events where the collected mass is larger than $\Delta_m M$ 
are defined as mergers. The value of $\Delta_m M$ is somewhat arbitrary,
but its physical meaning can be grasped in terms of modification of the
potential well of a cluster, following a merger. A major merger is expected
to appreciably change the dark matter distribution in the resulting cluster,
while only small perturbations are expected in small mergers, that are then
interpreted as events more similar to accretion than to real mergers.

Adopting this effective description of the merger and accretion events, it
is easy to use the LC formalism to construct simulated merger trees for
a cluster with fixed mass at the present time. Although useful from a 
computational point of view, this strategy of establishing a boundary 
between mergers and accretion events does not correspond to any
real physical difference between the two types of events, therefore in the 
following we will adopt the name ``merger'' for both regimes, provided there 
is no ambiguity or risk of confusion.

\begin{figure}[t!]
\plotone{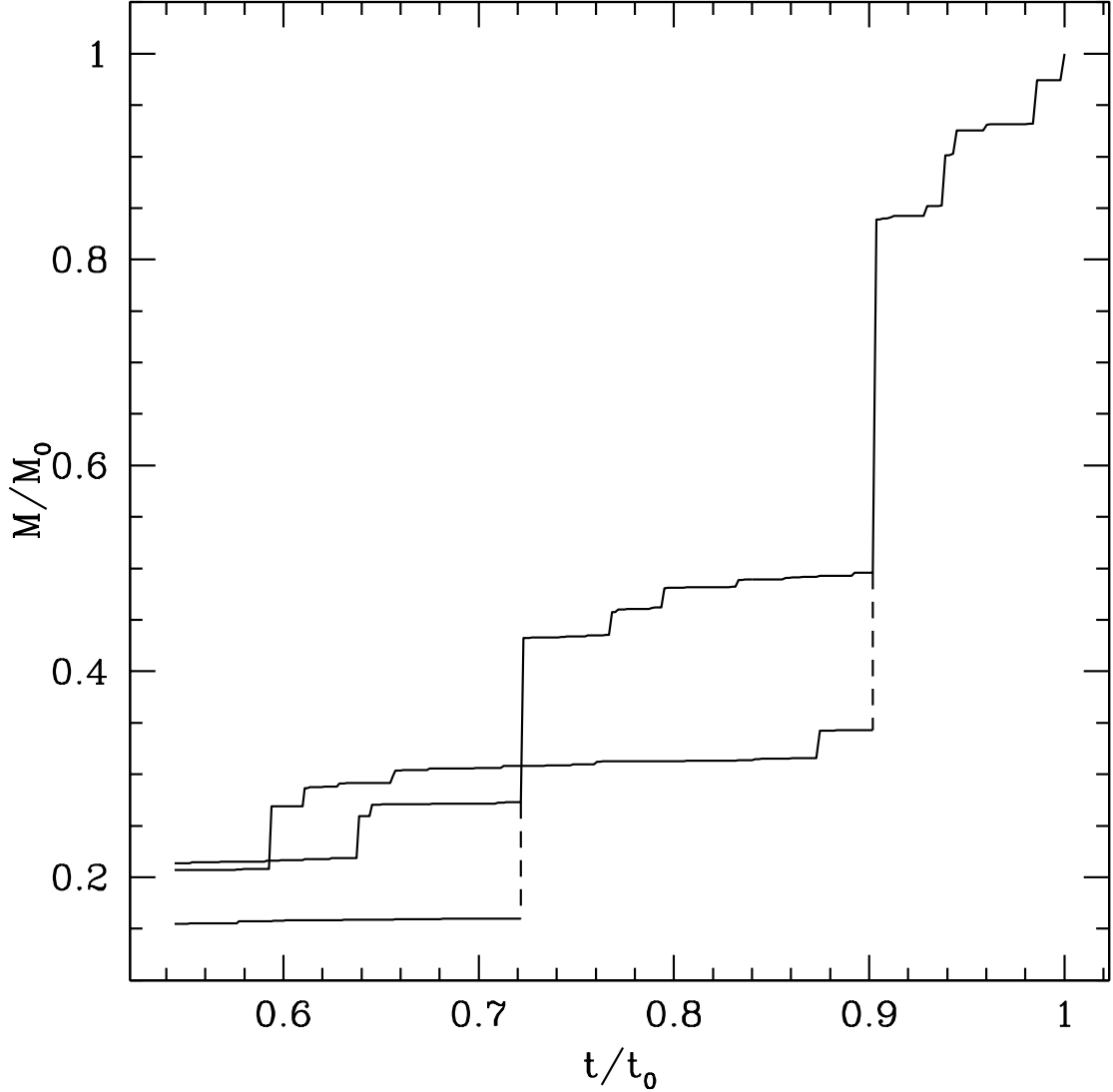}
\caption {Merger history of a cluster with present mass $10^{15}$ solar masses. The mass (y-axis) suffers major jumps in big merger
events. Time is on the x-axis.}
\label{fig:tree.eps}
\end{figure}

In fig. \ref{fig:tree.eps} we plotted a possible realization of the merger 
tree for a cluster with present mass equal to $10^{15} M_\odot$ and for 
$\Delta_m=0.6$. The history has been followed back in time up to redshift 
$z=3$. The big jumps in the 
cluster mass correspond to merger events, while smaller jumps
correspond to what Salvador-Sol\'e, Solanes \& Manrique (1998) defined as 
accretion events. 

%\placefigure{fig:mergertree} 
%Mshockfig1.ps

\section{Shocks during cluster mergers}\label{sec:shocks}

During the merger of two clusters of galaxies, the baryonic component,
feeling the gravitational potential created mainly by the dark matter 
component, is forced to move supersonically and shock waves are generated
in the intracluster medium.

In this section we describe in more detail the physical properties of such 
shocks, with special attention for their Mach numbers and compression 
factors. To this purpose we use an approach introduced in its original
version by Takizawa (1999). We assume to have two clusters, as completely 
virialized structures, at temperatures $T_1$ and $T_2$, and with masses 
$M_1$ and $M_2$ (here the masses are the total masses, dominated by the dark 
matter components). The virial radius of each cluster can be written as follows
$$
r_{vir,i} = \left(\frac{3 M_i}{4 \pi \Delta_c \Omega_m \rho_{cr} (1+z_{f,i})^3}
\right)^{\frac{1}{3}}=
$$
\begin{equation}
\left(\frac{G M_i}{100 \Omega_m H_0^2 (1+z_{f,i})^3}\right)^{\frac{1}{3}},
\label{eq:vir}
\end{equation}
where $i=1,2$, $\rho_{cr}=1.88 10^{-29} h^2 \rm{g}~\rm{cm}^{-3}$ 
is the current value of the critical density of the universe, 
$z_{f,i}$ is the redshift of formation of the cluster $i$, $\Delta_c=200$ is 
the density constrast for the formation of the cluster and $\Omega_m$ is the 
matter density fraction. 
In the right hand side of the equation we used the fact 
that $\rho_{cr}=3 H_0^2/8\pi G$, where $H_0$ is the Hubble constant.
The formation redshift $z_f$ is on average a decreasing function of the mass,
meaning that smaller clusters are formed at larger redshifts, consistently
with the hierarchical scenario of structure formation. There are intrinsic
fluctuations in the value of $z_f$ from cluster to cluster at fixed mass, 
due to the stochastic nature of the merger tree.

\begin{figure}[t!]
\plotone{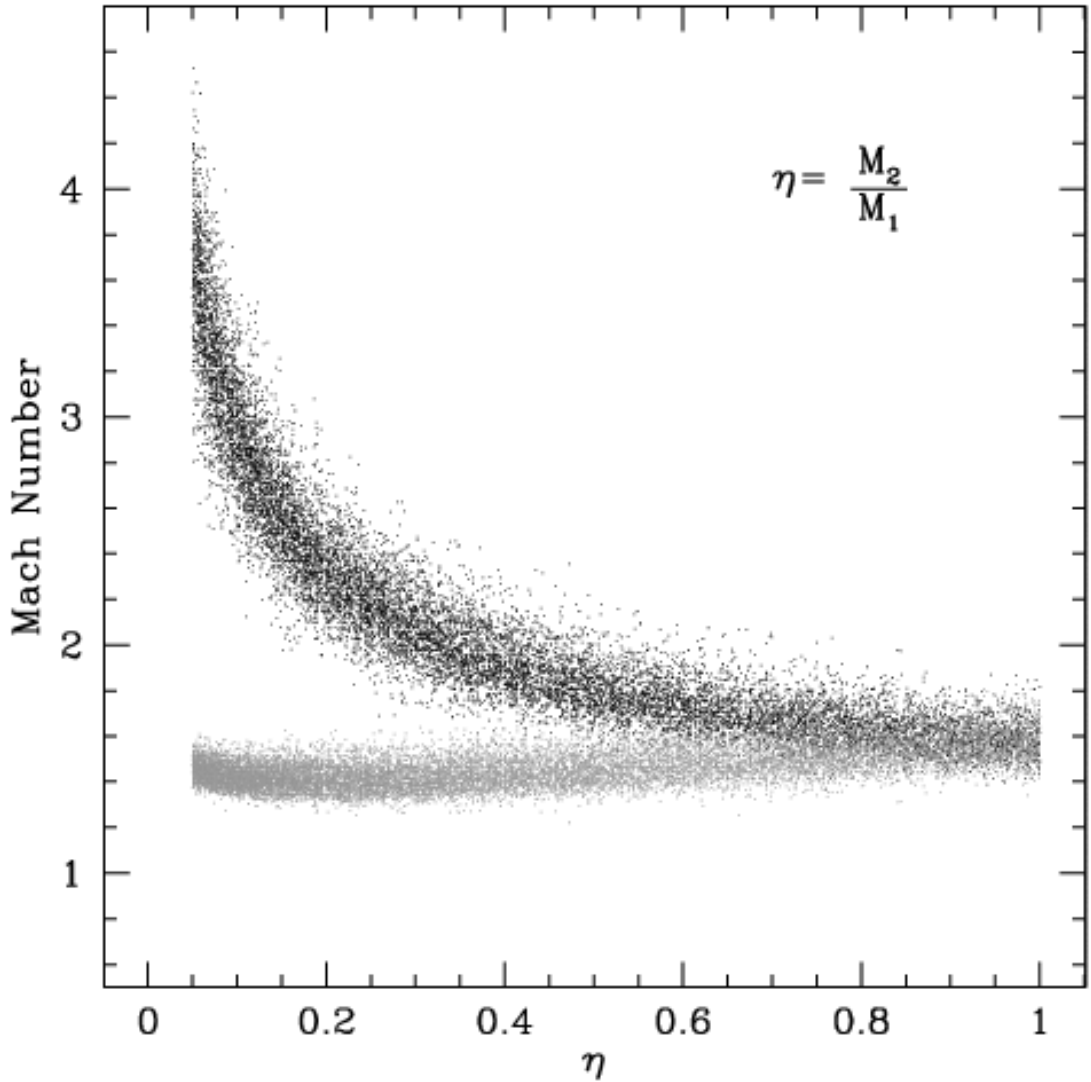}
\caption{Distribution of the Mach numbers of 
merger related shocks as a function of the mass ratio of the merging
subclusters. The upper strip is the distribution of Mach numbers in
the smaller cluster, while the lower strip refers to the bigger cluster.}
\label{fig:mach.ps}
\end{figure}

The relative velocity of the two merging clusters, $V_r$, can be easily 
calculated from energy conservation:
\begin{equation}
-\frac{G M_1 M_2}{r_{vir,1}+r_{vir,2}} + \frac{1}{2} M_r V_r^2 = 
-\frac{G M_1 M_2}{2 R_{12}},
\label{eq:Vrel}
\end{equation}
where $M_r=M_1 M_2 / (M_1+M_2)$ is the reduced mass and $R_{12}$ is the 
turnaround radius of the system, where the two subclusters are supposed to
have zero relative velocity. In fact the final value of the relative velocity
at the merger is quite insensitive to the exact initial condition of the two
subclusters. In an Einstein-De Sitter cosmology this spatial scale equals twice 
the virial radius of the system. Therefore, using eq. (\ref{eq:vir}), we get:
\begin{equation}
R_{12} = 2\,\left( \frac{M_1+M_2}{M_1}\right)^{1/3} \frac{1+z_{f,1}}{1+z_f} 
r_{vir,1}.
\end{equation}
where $z_f$ is the formation redshift of the cluster with mass $M_1+M_2$.
This expression remains valid in approximate way also for other cosmological
models (Lahav et al. 1991). The sound speed of the cluster $i$ is given by 
$$
c_{s,i}^2 = \gamma_g (\gamma_g - 1) \frac{G M_i}{2 r_{vir,i}},
$$
where we used the virial theorem to relate the gas temperature to the 
mass and virial radius of the cluster. The adiabatic index of the gas is
$\gamma_g=5/3$. The Mach number of each cluster while moving in the volume
of the other cluster can be written as follows:
\begin{eqnarray}
{\cal{M}}^2_1 & = & \frac{4(1+\eta)}{\gamma(\gamma-1)}
\left[\frac{1}{1+\frac{1+z_{f,1}}
{1+z_{f,2}}\eta^{1/3}}-\frac{1}{4\frac{1+z_{f,1}}{1+z_{f}}(1+\eta)^{1/3}}
\right] \nonumber \\
{\cal{M}}^2_2 & = & \eta^{-2/3}\frac{1+z_{f,1}}{1+z_{f,2}}{\cal{M}}^2_1, 
\end{eqnarray}
where $\eta=M_2/M_1<1$.
The procedure illustrated above can be applied to a generic couple of 
merging clusters, and in particular it can be applied to a generic merger
event in the history of a cluster with fixed mass at the present time.
The merger history (and indeed many realizations of the history) for a cluster 
can be simulated as discussed in the previous section. 
In particular, for a cluster with mass $M_0$ at the present time we simulate 
500 realizations of the merger tree and calculate the Mach numbers associated 
with the merger events. 
A value $\Delta_m=0.05$ is assumed. Note that this value is much 
lower than in (Fujita \& Sarazin 2001). This simply implies that we follow 
the histories of very small halos of dark matter, rather than the big ones 
only. The results of our calculations of the Mach numbers are plotted in 
fig. \ref{fig:mach.ps}. 

%\placefigure{fig:mach}
%Mshockfig2.ps

The striking feature of this plot is that for major mergers, involving 
clusters with comparable masses ($\eta\sim 1$), the Mach numbers of the 
shocks are of order unity. In other words the shocks are only moderately 
supersonic. 
In order to achieve Mach numbers of order of $3-4$ it is needed to consider 
mergers between clusters with very different masses ($\eta\sim 0.05$), which, 
in the language of Salvador-Sol\'e et al. (1998) and Fujita \& Sarazin (2001) 
would not be considered as mergers but rather as continuous accretion. These
events are the only ones that produce strong shocks, and this is of crucial 
importance for the acceleration of suprathermal particles, as discussed below. 
\begin{figure}[t!]
\plotone{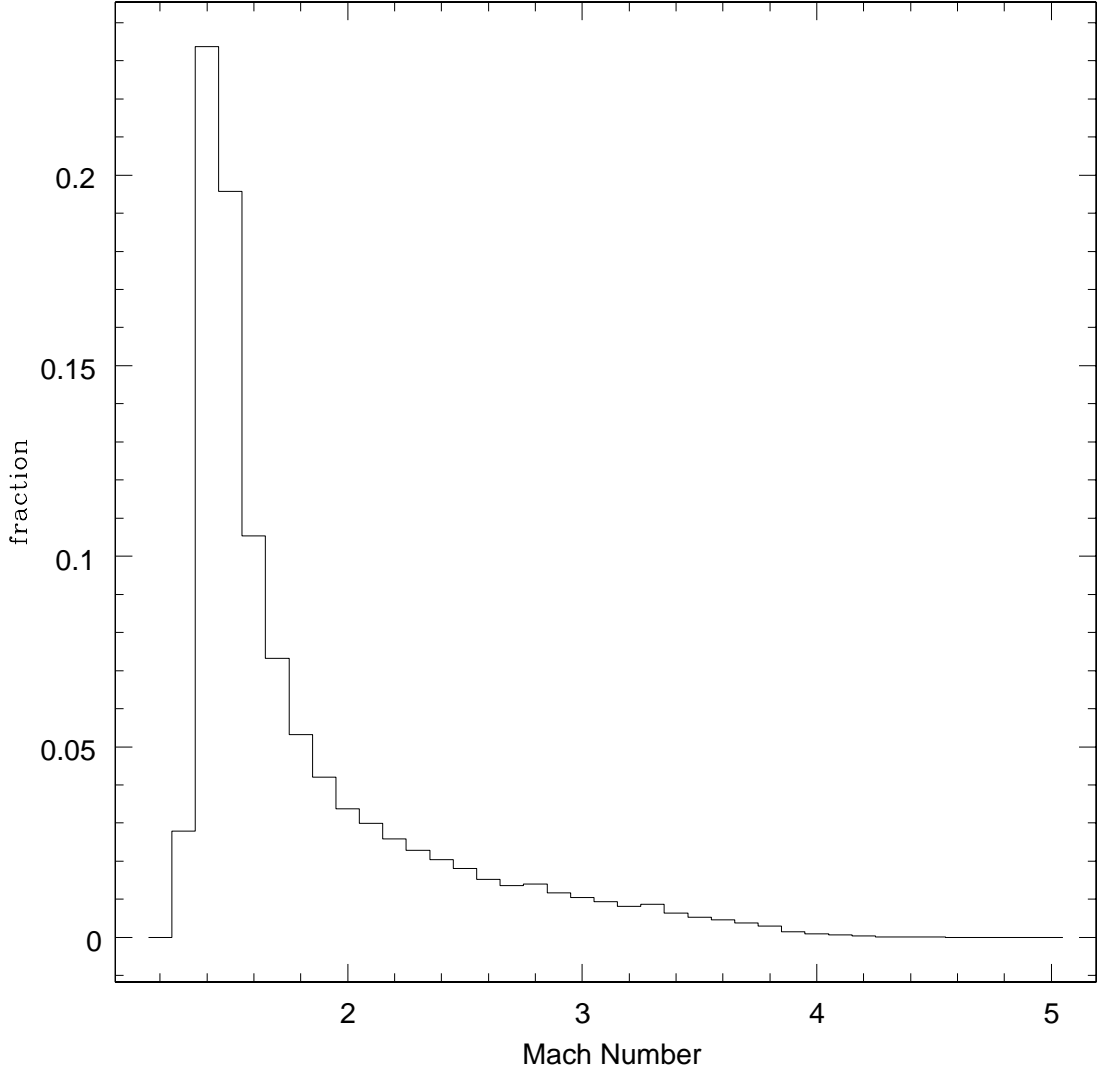}
\caption{Distribution of the Mach numbers of merger related shocks.}
\label{fig:istomach.ps}
\end{figure}
In fig. \ref{fig:istomach.ps} we plotted an histogram of the frequency of shocks
with given Mach number. It is easy to see that the greatest part of the merger
shocks are weak, with a peak in Mach number at about 1.4. This result is in some
contradiction with the Mach number distribution found by Miniati et al. (2000),
which is a bimodal distribution with one peak at Mach numbers of $\sim 1000$, and 
one at Mach numbers of $\sim 5$. While the former peak cannot be obtained within 
our approach, because it is caused by shocks in the cold outskirts of clusters,
not directly related to merger events, the latter shocks should be the same 
as those described in our paper. In fact, they are located, according to Miniati 
et al. (2000), within $0.5h^{-1}$ Mpc from the cluster center, namely in the 
virialized region, so that the arguments presented here do apply. It is worth
noticing in passing that the simulations of Miniati et al. (2000) have an 
intrinsic cutoff at Mach numbers smaller than $\sqrt 3$, so that the peak we
find at 1.5 is out of the available range. It is also worth noticing that 
the spatial resolution achieved in these simulations is of $0.315 h^{-1}$
Mpc, very close to the size of the region ($0.5h^{-1}$ Mpc) where the shocks
need to be identified.

On the other hand, the simple method introduced in our paper in order to
evaluate the Mach number of merger related shocks suffers of the limitation of
being applicable only to binary mergers. The cases in which a cluster is merging
with another cluster in a deeper gravitational potential well generated by a 
collection of nearby structures cannot be treated in the context of this method. 
The occurrence of these events may induce fluctuations around the mean Mach number
which might be larger than those considered here. We discuss this limitation more 
quantitatively in \S \ref{sec:results}.

The presence of the shocks at very large Mach numbers in the intergalactic medium
well outside virialized structures like clusters of galaxies can be expected simply 
on the basis of the propagation of shocks moving with speed typically $10^8 cm/s$
in a cold medium, with say $T\sim 10^4 K$. These shocks can accelerate
particles efficiently: if these particles are electrons, their radiation is
expected to be concentrated around the shocks, and well outside the intracluster
medium. If the accelerated particles are protons, they are expected to be 
advected inside the cluster by the accretion flow, where their inelastic 
collisions can generate secondary radiation (Gabici \& Blasi, in preparation).

Observationally, the strength of merger related shocks has been measured 
from high resolution X-ray images that allow one to evaluate the temperatures
on the two sides of a shock. One instance is provided by Cygnus A ($z=0.057$),
where two subclusters with comparable masses ($\eta\sim 1$) are merging.  
ASCA observations of this cluster have been used by Markevitch, Sarazin and
Vikhlinin (1999) to determine the temperature upstream ($T_1=4\pm 1$ keV)
and downstream ($T_2=8^{+2}_{-1}$ keV) of the shock, so that the compression
factor $r$ at the same shock can be determined from the Rankine-Hugoniot
relations:
\begin{equation}
\frac{1}{r} = \left[ 4\left(\frac{T_2}{T_1}-1\right)^2 + \frac{T_2}{T_1}
\right]^{1/2} - 2\left( \frac{T_2}{T_1} - 1\right).
\end{equation}
Numerically we obtain $r\sim 2.2-2.4$, corresponding to a Mach number 
${\cal M} \sim 2$, roughly consistent with our predictions in fig. 
\ref{fig:mach.ps}. Other cases of current merger events provide further 
examples of moderately supersonic relative motion of the merging clusters 
(Markevitch et al. 2001, 2002). 
None of the observed shocks seems to have Mach numbers around $\sim 5$, which 
appears in agreement with our findings. If the peak in Mach numbers were 
at $\sim 5$ it would be statistically unfavourable to observe only Mach
numbers smaller that this value.

\section{Acceleration and confinement of cosmic rays}\label{sec:accelera}

Relativistic particles can be accelerated at strong shocks by diffusive
(first order) Fermi acceleration (Fermi 1949; Blandford \& Eichler 1987).
This mechanism has been invoked several times as the ideal acceleration 
process in clusters of galaxies that are or have been involved in a merger
event (Blasi 2000; Sarazin \& Fujita 2001). In \S \ref{sec:fermi} we briefly 
summarize the basic physics of shock acceleration, since it is instrumental 
to understand whether merger related shock waves can indeed play a role for 
the acceleration of the relativistic particles responsible for the observed 
nonthermal radiation from clusters of galaxies. In \S \ref{sec:confinement} we 
summarize the findings of Berezinsky, Blasi \& Ptuskin (1997) on the 
confinement of cosmic rays in clusters of galaxies for cosmological time
scales. In \S \ref{sec:reacc} we briefly explain how to extend the formalism
of shock acceleration to the re-energization of the particles confined in
the cluster volume.

\subsection{Shock acceleration}\label{sec:fermi}

A shock with compression factor $r$ and Mach number ${\cal M}$ can accelerate
particles to a power law in momentum $f(p)\propto p^{-\alpha}$, with slope
$\alpha$ related to the Mach number and compression factor by the following
expressions:
\begin{equation} 
\alpha=\frac{r+2}{r-1} = 2 \frac{{\cal M}^2 + 1}{{\cal M}^2 - 1}.
\label{eq:slope}
\end{equation}
The acceleration occurs diffusively, in that particles scatter back and
forth across the shock, gaining at each crossing and recrossing an amount of
energy proportional to the energy of the particle itself, 
$\Delta E/E\sim V/c$, where $V$ is the speed of the shock and $c$ is the
speed of light.
The distribution function of the accelerated particles is normalized here 
by $\int_{p_{min}}^{p_{max}} dp E(p) f(p) = \eta_{shock}\rho u^2$, where $E(p)=
\sqrt{p^2 + m^2}$ and $m$ is the mass of the accelerated particles, 
$\eta_{shock}$ is an efficiency of acceleration, $\rho$ and $u$ are the 
density and speed respectively of the fluid crossing the shock surface. 
The minimum and maximum momenta ($p_{min}$ and $p_{max}$) of the accelerated 
particles are determined by the properties of the shock. In particular, 
$p_{max}$ is the result of the balance between the acceleration rate and either 
the energy loss rate or the rate of escape from the acceleration region. Less 
clear is how to evaluate $p_{min}$; the minimum momentum of the particles 
involved in the acceleration process depends on the microphysics of the shock,
which is very poorly known. Fortunately, most of the physical observables 
usually depend very weakly on $p_{min}$.  

The basic points introduced above can be simply applied to the case of 
shocks originated in cluster mergers, where the approximate values of the
parameters are known. In the following, we use these parameters to estimate 
the maximum energies attainable for electrons and protons as accelerated 
particles. The acceleration time, as a function of the particle energy $E$ 
can be written as
\begin{equation}
\tau_{acc} (E) = \frac{3}{u_1-u_2} D(E) \left[\frac{1}{u_1}+\frac{1}{u_2}
\right] = \frac{3 D(E)}{v^2} \frac{r(r+1)}{r-1},
\end{equation}
valid for any choice of the diffusion coefficient $D(E)$, for which we
consider two possible models. First we use the expression proposed in 
(Blasi \& Colafrancesco 1999): 
\begin{equation} 
D(E) = 2.3\times 10^{29} B_{\mu}^{-1/3} L_{20}^{2/3} E(GeV)^{1/3} cm^2/s,
\label{eq:dif1}
\end{equation}
where $B_{\mu}$ is the magnetic field in microgauss and $L_{20}$ is the 
largest scale in the magnetic field power spectrum in units of 20 kpc. 
Here we assumed that the magnetic field is described by a Kolmogorov power 
spectrum.

In this case the acceleration time becomes:
\begin{equation}
\tau_{acc} (E) \approx 6.9\times 10^{13} B_{\mu}^{-1/3} L_{20}^{2/3} 
E(GeV)^{1/3} v_8^{-2} g(r)~~~ s,
\end{equation}
where $v_8=\frac{v}{10^8 cm/s}$ and $g(r)=r(r+1)/(r-1)$ and $v=u_1$.

For electrons, energy losses are dominated by ICS off the photons of the 
cosmic microwave background, provided the magnetic field is smaller than
$\sim 3 \mu G$. The maximum energy of accelerated electrons is obtained by 
requiring $\tau_{acc}<\tau_{loss}$:
\begin{equation}
E_{max}^e \approx 118 L_{20}^{-1/2} B_{\mu}^{1/4}  
v_8^{3/2} g(r)^{-3/4}~ GeV.
\end{equation}
The relation between the compression ratio $r$ and the Mach number is
\begin{equation}
r=\frac{\frac{8}{3} {\cal M}^2}{\frac{2}{3} {\cal M}^2 + 2},
\end{equation}
valid for an ideal monoatomic gas. Here ${\cal M}$ is the Mach number of
the unshocked gas, moving with speed $v_8$. 

For protons, energy losses are usually not relevant and the maximum energy 
is clearly determined by the finite time duration of the merger event. 
Therefore the maximum energy for protons will be defined by the condition 
$\tau_{acc}<t_{merger}$, which gives 
\begin{equation}
E_{max}^p \approx 9\times 10^7 L_{20}^{-2} B_{\mu} v_8^6 g(r)^{-1/2}~GeV,
\end{equation}
for $t_{merger}\sim 10^9$ years.
As a second possibility for the diffusion coefficient we assume Bohm diffusion,
well motivated for the case of strong turbulence. In this case:
\begin{equation}
D(E)=3.3\times 10^{22} E/B_\mu ~ cm^2/s.
\label{eq:dif2}
\end{equation}
For electrons we obtain:
\begin{equation}
E_{max}^e \approx 6.3\times 10^4 B_{\mu}^{1/2} v_8 g(r)^{-1/2}~ GeV,
\end{equation}
while for protons
\begin{equation}
E_{max}^p \approx 3\times 10^9 B_{\mu} v_8^2 g(r)^{-1}~GeV.  
\end{equation}
If $E_{max}^p$ becomes larger than $\sim 10^{10}$ GeV, energy losses due
to proton pair production and photopion production on the photons of the microwave
background become important and limit the maximum energy to less that 
a few $10^{10}$ GeV (Kang, Rachen \& Biermann 1997).

The relative abundance of electrons and protons at injection is an unknown 
quantity. 
Theoretically, there are plausibility arguments for having a small $e/p$ 
ratio, related to the microphysics of shock acceleration: protons resonate with 
Alfv\'en waves on a wide range of momenta, so that they can be efficiently 
extracted from the thermal distribution and injected into the acceleration 
engine. For electrons this is much harder. Low energy electrons do not interact 
with Alfv\'en waves and some other modes need to be excited (for instance 
whistlers) and sustained against their strong damping.
A detailed discussion of the electron injection at non relativistic shocks 
can be found in (McClements et al. 1997; Levinson 1994) and references therein.

Another issue that contributes to the suppression of the injection of the 
electron component in a shock is the finite thickness of the shock, comparable
with the Larmor radius of thermal protons (Bell 1978a,b). Electrons can be 
injected in the shock accelerator only if their Larmor radius is larger than
the thickness of the shock. For a proton temperature of $\sim 8$ keV, only 
electrons with energy larger than $\sim 5-10$ MeV can be injected in the 
acceleration box. This energy is much larger than the typical electron 
temperature in the ICM. The value of $5$ MeV can be adopted as a sort of low 
energy cutoff in the injection spectrum of electrons.

\subsection{Confinement of accelerated particles}\label{sec:confinement}
 
It was first realized by Berezinsky, Blasi \& Ptuskin (1997) and 
Volk, Aharonian, \& Breitschwerdt (1996) that the bulk of the hadronic
cosmic rays accelerated or injected in clusters of galaxies remain 
diffusively confined within the cluster volume for cosmological times.
The maximum energy for which the confinement is effective is a strong
function of the assumed diffusion coefficient, being the highest for
the case of Bohm diffusion. For this case, for our purposes we can 
assume a complete confinement. In cases that might be more realistic,
for instance for a Kolmogorov spectrum of magnetic fluctuations, the 
confinement may be limited to particles with energy up to a few TeV.

For the energies at which the confinement is effective, the energy 
density in cosmic rays increases with time due to the pile up of 
particles injected during the cluster history. In particular this
accumulation occurs at each merger event if the merger related shocks 
are strong enough. This process may result in a sufficient accumulation 
of cosmic ray protons to induce detectable fluxes of gamma radiation 
through pion decay (see Blasi (2002) for a recent review).

For high energy electrons, the processes that affect the most the propagation 
are energy losses, so that the discussion on confinement does not apply.
On the other hand, the confined protons induce a secondary population of 
electron-positron pairs due to the decay of charged pions generated in
inelastic collisions of relativistic protons with the thermal gas.
These pairs are responsible for radio radiation through synchrotron 
emission and for X-rays due to ICS. Moreover, gamma radiation is also
generated through ICS and bremsstrahlung emission of these secondary pairs.

Low energy electrons ($\gamma\sim 100$) lose energy slowly and can be piled 
up and confined for a few billion years, and eventually re-energized by shocks
related to successive merger events. 

\subsection{Reacceleration of confined relativistic particles}\label{sec:reacc}

Shocks formed during a merger not only accelerate new particles from the
thermal bath but also re-energize particles confined in the cluster either 
from previous mergers or from additional sources of cosmic rays. Low energy
electrons and relativistic hadrons are affected by the reacceleration.

Assuming that the spectrum of particles confined within the cluster is 
$n(p)$, and that the merger shock has a compression factor $r$, the spectrum 
of the reaccelerated particles can be easily calculated. In fact, particles 
with initial momentum between $p$ and $p+dp$ are $n(p)dp$ and they are 
reprocessed by the shock into the spectrum $df(p')$ that can be calculated 
by standard shock acceleration theory 
\begin{equation}
df(p') = dA \left(\frac{p'}{p}\right)^{-\alpha} ~~~ \alpha=\frac{r+2}{r-1} 
\end{equation}
and
\begin{equation}
\int_p^{\infty} dp' df(p') = n(p) dp \to dA = (\alpha-1) n(p) \frac{dp}{p},
\end{equation}
where we assumed $\alpha>1$ and $p_{max}\to \infty$. The spectrum of all 
the reaccelerated particles is then:
\begin{equation}
f(p') = \int_{p_{min}}^{p'} \frac{dp}{p} (\alpha-1) n(p)  
\left(\frac{p'}{p}\right)^{-\alpha}.
\label{eq:reacc}
\end{equation}
If the initial spectrum of the particles is a power law 
$n(p)=n_0 (p/p_{min})^{-\beta}$, then eq. (\ref{eq:reacc}) gives
\begin{equation}
f(p') = \frac{\alpha-1}{\alpha-\beta} n_0 
\left(\frac{p'}{p_{min}}\right)^{-\alpha} \left[ 
 \left(\frac{p'}{p_{min}}\right)^{\alpha-\beta} - 1\right]
\end{equation}
for $\alpha>\beta$ (a similar expression is obtained for $\beta>\alpha$), and
\begin{equation}
f(p') = (\alpha-1) n_0 
\left(\frac{p'}{p_{min}}\right)^{-\alpha} \ln 
\left(\frac{p'}{p_{min}}\right)
\end{equation}
for $\alpha=\beta$. A typical effect of reacceleration is to generate particle
spectra which are flatter than the spectrum of the seed particles. Moreover, at
each reacceleration step, the total energy in accelerated particles can
be substantially increased if the shocks are sufficiently strong. 
Note that eq. (\ref{eq:reacc}) holds for an arbitrary spectrum of the 
pre-existing particles.

Low energy electrons, which may also be confined for a few billion years
with no appreciable energy loss within the clusters, may also be re-energized 
by the passage of merger shocks, so that for the duration of the merger they 
can reach the relativistic energies required to produce the nonthermal radiation 
observed in the radio and X-ray bands. Strong shocks, with Mach numbers of order 
$\sim 3-4$ are needed in order to explain observations.

\section{Simulated shocks and spectra of nonthermal particles}\label{sec:results}

The technique described in \S \ref{sec:simula} allows us to simulate merger
trees of a cluster and evaluate the physical properties of the shock waves 
generated at each merger of subclusters, as discussed in \S \ref{sec:shocks}.
While the merger tree is constructed moving backwards in time, starting with
a cluster of given mass at the present time, the shock properties and particle
acceleration are reconstructed by moving forward in time and accounting for the 
mergers of all the substructures generated during the first step of the simulation
(backwards in time). 

Each merger here is assumed to be a two body event. This is clearly increasingly 
less true for large mass differences between the merging subclusters, because it 
may happen  that more subclusters with small masses can {\it accrete} onto the big 
cluster at approximately the same time. In other words, the merger between two 
clusters may occur in the deeper gravitational potential well created by nearby 
structures. In this case, the relative velocity between the two clusters, and also 
the related shock Mach numbers  may be larger (or smaller) than those estimated in 
\S \ref{sec:shocks}. Although a rigorous evaluation of the probability of occurrence
of this kind of situations cannot be carried out in the context of our simple 
approach, convincing arguments can be provided to support the results discussed in 
\S \ref{sec:shocks}: let us assume that our two clusters, with mass $M_1$ and 
$M_2$, are merging in a volume of average size $R_{sm}$ where the overdensity if 
$1+\delta$ ($\delta=0$ corresponds to matter density equal to the mean value). 
Clearly the overdense region must contain more mass than that associated with the 
two clusters, therefore for a top-hat overdensity at $z=0$ we can write:
\begin{equation}
\frac{4}{3} \pi R_{sm}^3 \rho_{cr} \Omega_m (1+\delta) = \xi (M_1+M_2),
\end{equation}
where $\xi>1$ is a measure of the mass in the overdense region in excess of 
$M_1+M_2$. In numbers, using $\Omega_m=0.3$, this condition becomes:
\begin{equation}
(1+\delta) = 2 \xi M_{15} R_{10}^{-3},
\label{eq:delta}
\end{equation}
where $M_{15}$ is $M_1+M_2$ in units of $10^{15}$ solar masses and
$R_{sm}=10~{\rm Mpc}~R_{10}~h^{-1}$. 

If the clusters are affected by the potential well of an overdense region with
total mass $M_{tot}$, the maximum relative speed that they can acquire is 
$v_{max}\approx 2 \sqrt{G M_{tot}/R_{sm}}$. Note that this would be the relative 
speed of the two clusters if they merged at the center of the overdense region and 
with a head-on collision, therefore any other (more likely) configuration would 
imply a relative velocity smaller than $v_{max}$. In particular, the presence of
the local overdensity might even cause a slow down of the two merging clusters,
rather than a larger relative velocity. 
In numbers 
$$ v_{max} =1.1\times 10^8 \xi^{1/2} M_{15}^{1/2} R_{10}^{-1/2} ~
{\rm cm/s}.$$
Using the usual expression for the sound speed in a cluster with mass $M_i$ we also
get
$$c_s = 8.8 \times 10^7 M_{i,15}^{1/3}~ {\rm cm/s}.$$
Therefore the maximum Mach number that can be achieved in the i-th cluster is
\begin{equation}
{\cal M}_{i,max} = 1.25 \xi^{1/2} R_{10}^{-1/2} M_{15}^{1/2} M_{i,15}^{-1/3}.
\end{equation}
As stressed in the previous sections, the Mach numbers which may be relevant for
particle acceleration are ${\cal M}>3$, which implies the following condition on 
$\xi$:
\begin{equation}
\xi>5.8 M_{15}^{-1} M_{i,15}^{2/3}R_{10},
\end{equation}
that, when introduced in eq. (\ref{eq:delta}) gives:
\begin{equation}
(1+\delta) > 11.6 R_{10}^{-2} M_{i,15}^{2/3}.
\label{eq:overdensity}
\end{equation}
Similar results may be obtained using the velocity distribution of dark matter 
halos as calculated in semi-analytical models (Sheth \& Diaferio 2001) and 
transforming this distribution into a pairwise velocity distribution, by adopting
a suitable recipe.

The probability to have an overdensity $1+\delta$ in a region of size $R_{sm}$
has the functional shape of a log-normal distribution, as calculated by Kayo, 
Taruya \& Suto (2001). Eq. (\ref{eq:overdensity}) gives the overdensity 
$1+\delta$ necessary for a cluster of mass $M_i$ to achieve a Mach number at 
least 3 in the collision with another cluster in the same overdense region. 
In the table below we report the probabilities $P(\delta)$ as a function 
of $M_i$, evaluated following Kayo et al. (2001) for different sizes of the 
overdense region. 
\begin{center}
\begin{tabular}{|c|c|c|c|} \hline
$M_{i,15}$ & $R_{10}$ & $(1+ \delta)$ & $P(\delta)$ \\ \hline
1   & 0.8 & 18.1 & $7\times 10^{-5}$\\ 
1   & 1   & 11.6 & $9\times 10^{-5}$\\ 
1   & 1.2 & 8.1 & $10^{-4}$\\
0.5 & 0.8 & 11.4 & $5\times 10^{-4}$\\
0.5 & 1   & 7.3 & $10^{-3}$\\
0.5 & 1.2 & 5.1 & $2\times 10^{-3}$\\
0.1 & 0.8 & 3.9 & $2\times 10^{-2}$\\
0.1 & 1   & 2.5 & $5\times 10^{-2}$\\
0.1 & 1.2 & 1.7 & 0.12\\ \hline
\end{tabular}
\end{center}
These numbers must be interpreted as upper limits to the probability that 
a cluster of mass $M_i$ develops a merger shock with Mach number larger than three,
since the probabilities refer to the configuration in which the relative velocity 
between the two clusters is maximized. For rich clusters, with masses larger than 
$5\times 10^{14}$ solar masses (corresponding to X-ray luminosities 
$L_X>4\times 10^{44} erg/s$) 
the probability that the presence of a local overdensity may generate Mach numbers
relevant for the nonthermal activity has been estimated to be $\sim 10^{-3}$ or
smaller, suggesting that our two body approximation is reasonable, in particular
for the massive clusters that are typically observed to have nonthermal activity. 
For smaller clusters, the probabilities become higher, indicating that the 
distribution of Mach numbers might have a larger spread compared with that 
illustrated in fig. 2. 
Note however that for small clusters, even the two body approximation gives 
relatively high Mach numbers, provided the merger occurs with a bigger cluster,
simply as a result of a lower temperature and a correspondingly lower sound speed. 

Motivated by this argument, in the following we will continue to consider only
binary mergers.

%Although this would slightly affect the
%relative merger velocity, it does not change in a significant way our
%predictions of the expected Mach numbers. 

For each of these binary mergers, we evaluate the duration of the merger and 
the relative volumes that take part in it.

\begin{figure}[t!]
\plotone{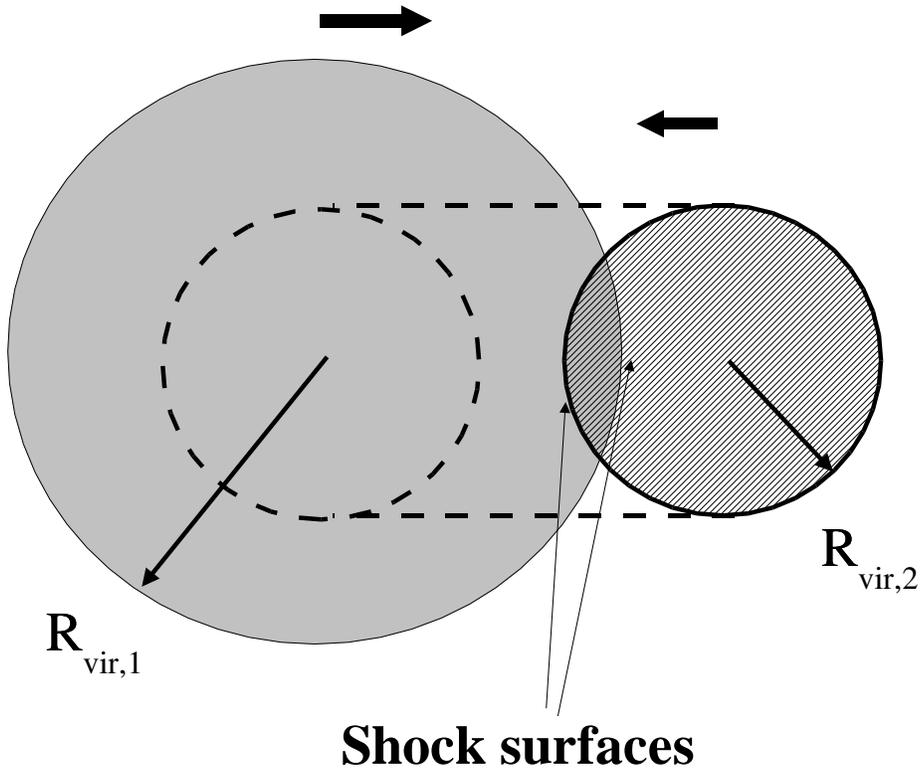}
\caption{Geometry of the merger between two subclusters.}
\label{fig:geometry.eps}
\end{figure}

A prescription on how to define these two quantities is required since they 
affect the amount of energy injected in the cluster in the form of relativistic
particles. Acceleration occurs at each one of the two shocks formed as a result
of the merger. Once the relative velocity is calculated from eq. 
(\ref{eq:Vrel}), the duration of the merger event is defined as 
$$
\tau_{mer} \approx \frac{r_{vir,1}}{V_r}
$$
where $r_{vir,1}$ is defined as the virial radius of the bigger cluster.
The geometry of the merger is schematized in fig. \ref{fig:geometry.eps}.

It is extremely important to keep in mind that the radiation from electrons 
directly accelerated at a shock is, in first approximation, solely dominated 
by the last merger events that the cluster suffered; on the other hand, for 
protons and the consequent secondary products, it is mandatory to study the 
full history of the cluster and account for all the shocks that traversed the 
cluster volume. Both acceleration of new particles and re-energization of 
previously accelerated particles need to be taken into account, as explained 
in \S \ref{sec:accelera}.

Observations may be explained by a population of electrons (primaries or 
secondaries) injected in the cluster with a power law spectrum having a slope 
around $2.3-2.4$ (see for instance (Blasi 2000)). In the discussion below
we will refer exclusively to injection spectra, without explicitely accounting
for the obvious steepening by one power in energy in the equilibrium spectrum
of the radiating electrons (energy losses are dominated by ICS and synchrotron
emission in the energy range of interest). In other words, if the injection 
spectrum of electrons (as primaries or as secondary products of hadronic 
interactions) is (locally) a power law with slope $2.3-2.4$, the corresponding 
equilibrium spectra will have local slope $3.3-3.4$. 

\begin{figure}[t!]
\plotone{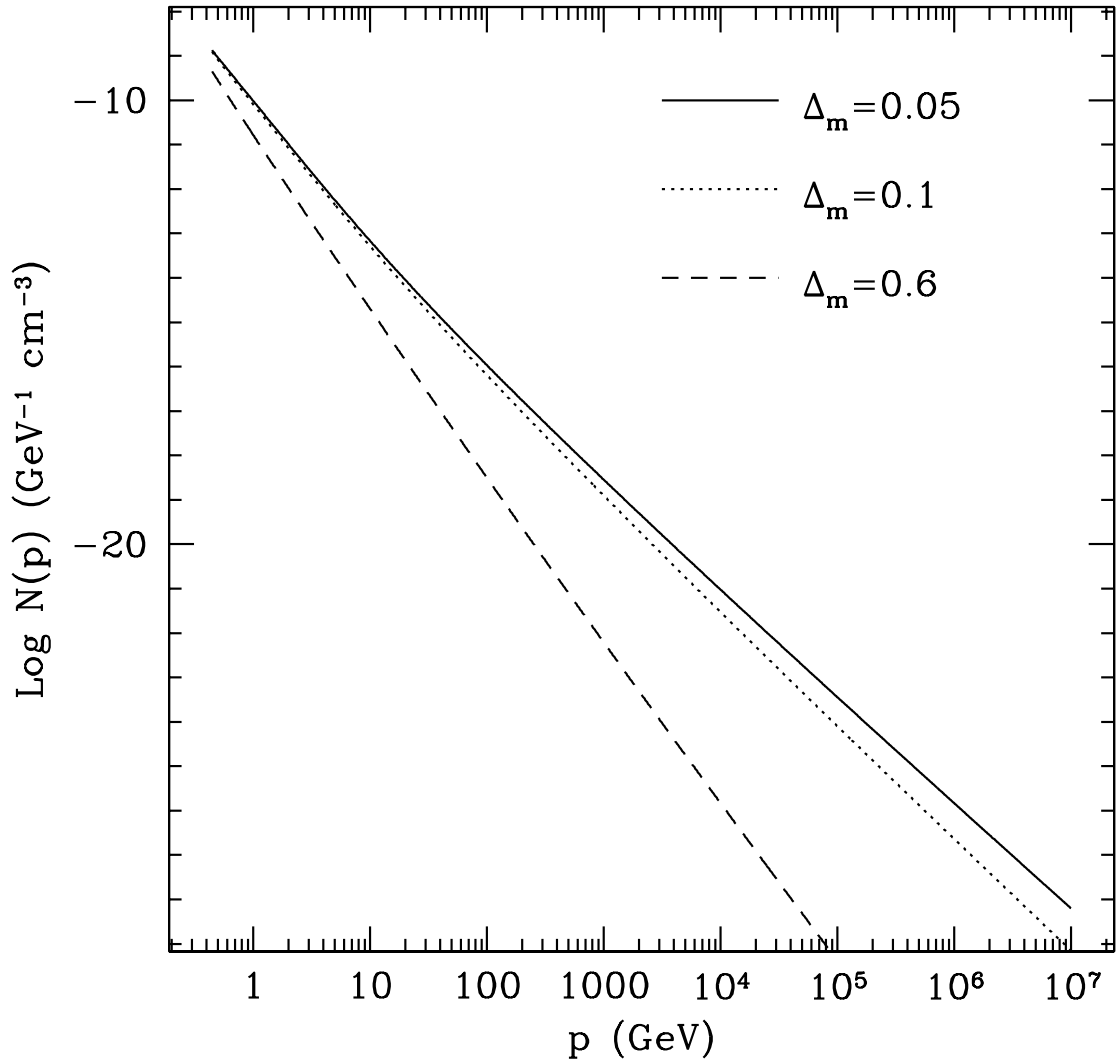}
\caption{Time-integrated average proton spectra resulting from all the mergers in 
a cluster. The three curves are obtained for $\Delta_m=0.05$ (solid line), 
$\Delta_m=0.1$ (dotted line) and $\Delta_m=0.6$ (dashed line).}
\label{fig:spectra.ps}
\end{figure}

In the following we distinguish the two cases of primary electrons (directly 
accelerated at the shocks) and secondary electrons, whose spectrum at energies
above a few GeV approximately reproduces the spectrum of the parent protons. 
We start with the case of secondary electrons, concentrating our attention 
upon the spectrum of the accelerated protons. We stress again that merger 
related shocks accelerate {\it new}  protons and reaccelerate protons which 
were already confined in the parent clusters from previous times. The 
efficiency for particle acceleration at each shock is taken as a constant 
equal to $10\%$. For simplicity we assume that the accelerated particles are 
all confined in the cluster volume, although this may not be true at the 
highest energies for large diffusion coefficients. It seems however a good 
approximation for protons with the energies that we are interested in. 
The spectra that result from our calculations are plotted in fig. 
\ref{fig:spectra.ps}. These curves are obtained averaging the spectra of 
500 clusters, each one followed in its merger history back in time to redshift
$z=3$. We consider the three cases $\Delta_m=0.6$ (dashed line), $\Delta_m=0.1$ 
(dotted line) and $\Delta_m=0.05$ (solid line).
It is clear that at high energy it is crucial to account for small mergers,
since they are responsible for flatter spectra. In fact the flatter regions
are also due to some level of reacceleration of pre-existing protons, confined
in the intracluster volume. If only major mergers are considered 
($\Delta_m=0.6$), the resulting spectrum is too steep to be of any relevance 
for the generation of the observed nonthermal radiation.

\begin{figure}[t!]
\plotone{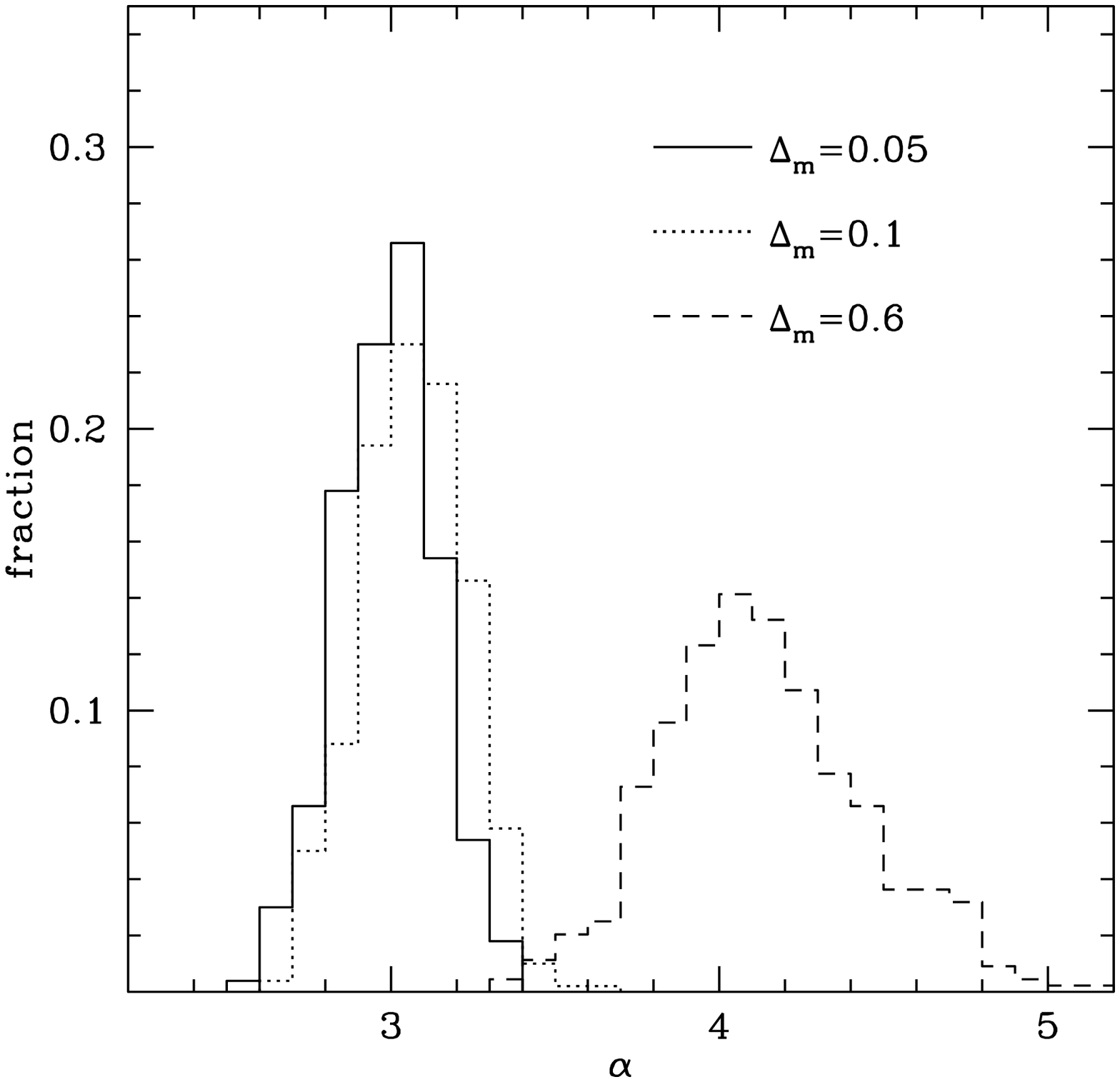}
\caption{Slope of the time-integrated  proton
spectra resulting from all the mergers in a cluster. The three curves
are obtained for $\Delta_m=0.05$ (solid line), $\Delta_m=0.1$ (dotted line)
and $\Delta_m=0.6$ (dashed line)}
\label{fig:slopes.ps}
\end{figure}
In fig. \ref{fig:slopes.ps} we plot the distribution of spectral slopes
($\alpha$ in eq. (\ref{eq:slope})) for different choices of the threshold
in mass ratio for mergers. Since the spectra are not power laws (see 
fig. \ref{fig:spectra.ps}) it is convenient to plot the slopes at fixed energy,
say 10 GeV (it is worth to recall that protons with this energy typically
generate electrons of few GeV, which are the ones relevant for the production 
of radio halos). When major mergers are considered ($\Delta_m=0.6$), the typical 
slopes of the spectra of accelerated particles peak around 4.4, 
too steep to be relevant for nonthermal radiation in clusters.
When mergers between clusters with very different masses are considered, 
the situation improves, but still, even for $\Delta_m=0.05$, the spectra
remain too steep. An important point is that flat spectra, if any, are not
obtained in major mergers but rather in those mergers that according to 
Fujita \& Sarazin (2001) qualify as {\it accretion} events. Shocks related
to major mergers are not strong enough to account for the observed spectra.
The spread in the values of $\alpha$ in fig. \ref{fig:slopes.ps} reflects the
fluctuations in the Mach numbers in fig. \ref{fig:mach.ps}, due to the
distribution of formation redshifts in the simulation and to the stochasticity 
of the merger tree.
Even accounting for these fluctuations, the strength of the shocks appears 
to be insufficient to generate the required spectra of accelerated particles. 

\begin{figure}[t!]
\plotone{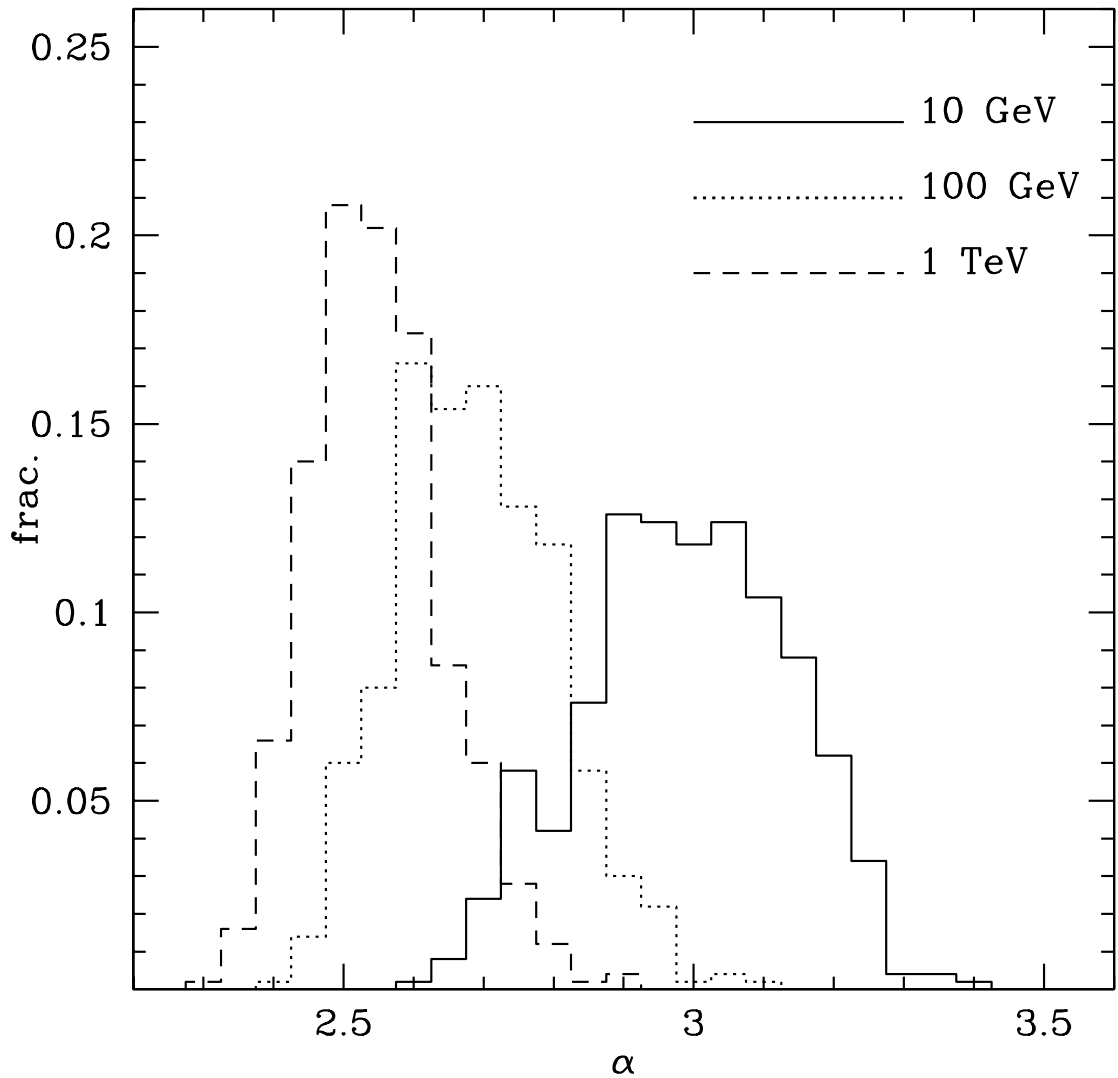}
\caption{Slope of the time-integrated proton spectra resulting from all the 
mergers in a cluster for $\Delta_m=0.05$ at different proton energies: 10 GeV 
(solid line), 100 GeV (dotted line) and 1 TeV (dashed line).}
\label{fig:slopeEdep.ps}
\end{figure}

In fig. \ref{fig:slopeEdep.ps} we also plot the slope of the proton spectrum 
for $\Delta_m=0.05$, at three energies, 10 GeV (solid line), 100 GeV (dotted
line) and 1 TeV (dashed line). It is clear that the spectrum flattens at high
energies, which is again in some disagrement with observations of radio halos,
which seem to suggest a steepening of the radio spectrum towards its higher
frequency end.
We now consider the acceleration of primary electrons as responsible for 
nonthermal radiation in clusters of galaxies. In this case, only recent
mergers, occurred within about one billion years may be related to the 
observed radiation, due to the short lifetimes of relativistic electrons. 
Therefore, we simulate the merger tree of clusters with fixed present mass, 
limiting the simulations to one billion years far into the past. In fact, even 
one billion years is a time appreciably longer than the time for losses of
relativistic electrons, so that our predictions have to be considered as
optimistic. Mergers and related shocks have been treated as discussed in the 
previous sections. 

\begin{figure}[t!]
\plotone{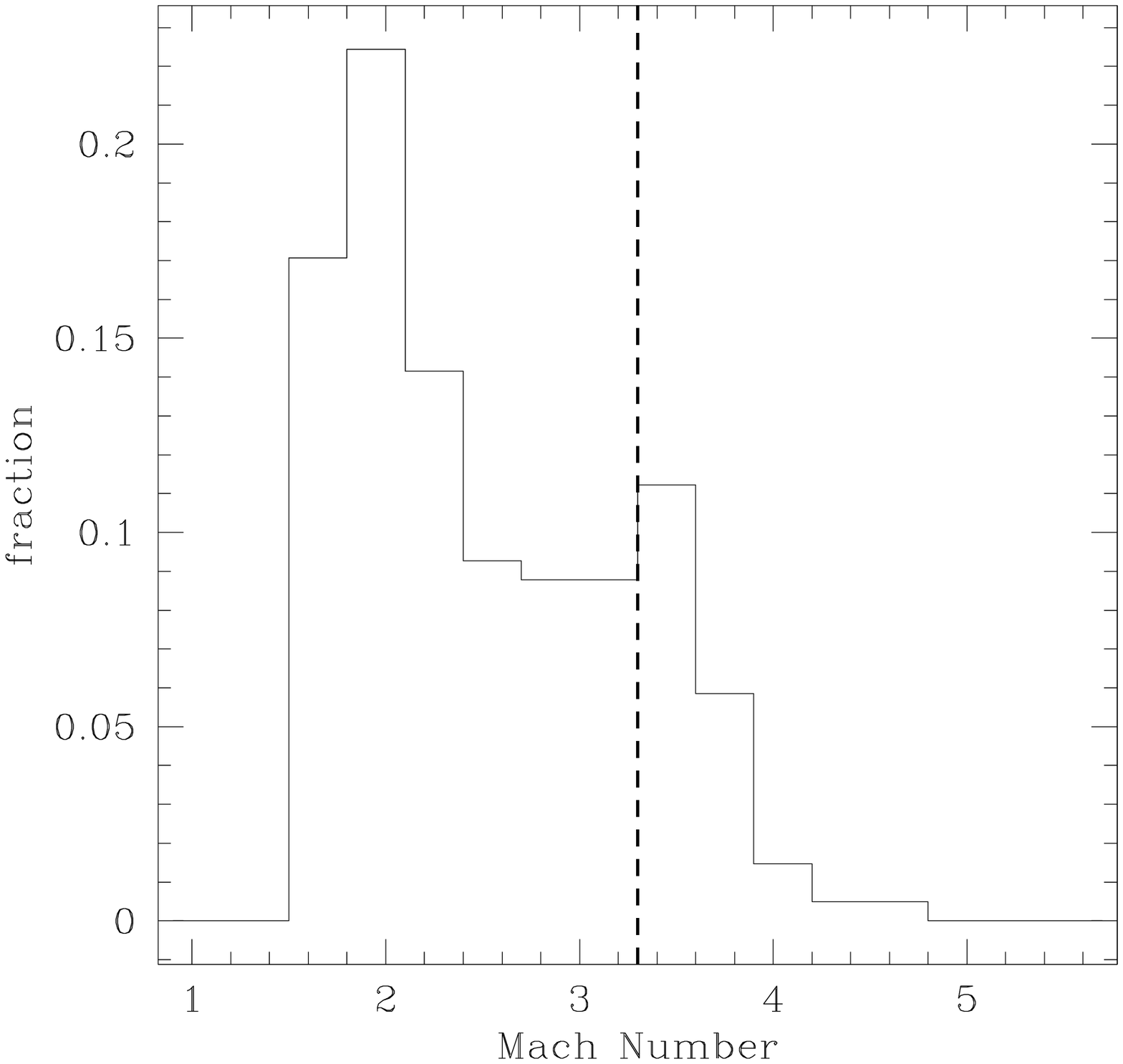}
\caption{Distribution of Mach numbers for the mergers occurred in the last
one billion years for a cluster with present mass $10^{15}$ solar masses. The 
dashed line indicates the Mach number that corresponds to shocks able to accelerate
electrons with a spectrum $E^{-2.4}$. The histogram is obtained averaging over 500
clusters.}
\label{fig:primary.ps}
\end{figure}

Of the 500 clusters with mass $10^{15}$ solar masses that we simulated the
merger tree of, about $30\%$ suffered a merger during the last billion years.
In fig. \ref{fig:primary.ps} we plotted the Mach numbers of the shocks generated
in the clusters that suffered at least one merger (actually it is rare to have
more than one merger during the last billion years). The dashed line represents 
the value of the Mach number that would correspond to a slope of the accelerated
electrons equal to 2.4, required to explain observations. Only $20\%$ of the 
shocks have Mach numbers fulfilling this condition, so that in the end, about 
$6\%$ of the 500 simulated $10^{15}$ solar masses clusters of galaxies have
suffered a merger than may have generated nonthermal activity with Coma-like
spectrum.

\section{Conclusions}\label{sec:conclude}

We investigated the possibility that the nonthermal activity observed from
some clusters of galaxies may originate through radiative losses of electrons
either accelerated at shocks during cluster mergers or produced as secondary 
products of the inelastic collisions of protons, in turn accelerated at the 
same merger shocks.

While the spectrum of protons at any time is the result of all the merger history 
of a cluster, due to cosmic ray confinement (Berezinsky, Blasi \& Ptuskin 1997;
Volk, Aharonian \& Breitschwerdt 1996), primary electrons that are able to radiate
radio or X-ray photons at present need to be accelerated in very recent times, so
that only the last mergers are relevant (Fujita \& Sarazin 2001). 

The merger history of clusters can be simulated by using a PS approach, which 
allows one to obtain, for each merger event, the mass of the subclusters. In the
reasonable assumption of binary mergers, it is also easy to calculate the 
relative velocity of the two merging clusters, and if clusters are assumed to 
be virialized structures, also the Mach numbers of the two approaching subclusters.
Once the Mach numbers are known, it is possible to calculate the spectra of the
particles accelerated by first order Fermi acceleration. 

We discussed separately the case of secondary and primary electrons. The secondary 
electrons, generated in $pp$ scattering have a spectrum that approximately
reproduces the spectrum of the parent protons (due to Feynman scaling for the cross
section). The spectrum of the protons is calculated by taking into account the
acceleration process at each merger, and the reacceleration of protons confined
in the merging clusters. The spectrum at the present time is the result of these
processes along the all history of the cluster. The spectra that we obtained from
our calculations are typically steeper than those required to explain the observed
nonthermal radiation. This result is the consequence of the weakness of the shocks
associated to major mergers, where the relative motion of the two subclusters
occurs at almost free-fall velocity, which implies Mach numbers only slightly 
larger than unity. Minor mergers produce stronger shocks, but they are 
energetically subdominant. 

Mach numbers larger than those calculated here may be achieved if the merger
event occurs in an overdense region, where the infall motion of the two clusters
may be dominated by the gravitational well of the surrounding matter rather than
by the mutual interaction between the two clusters. This kind of situations can 
indeed either increase or decrease the Mach numbers compared with the binary 
case. In \S \ref{sec:results} we have found that in order to obtain Mach numbers
larger than $\sim 3$ the overdensity must be such that for rich clusters (mass
larger than $5\times 10^{14}$ solar masses) the probability of sitting in such
a potential is pretty slim, and the binary merger model should represent an
accurate description of reality, in a statistical sense.

As stressed above, primary electrons can generate observable nonthermal radiation
only if accelerated in recent mergers, occurred less than $10^9$ years ago. Our
simulations of 500 clusters with mass $10^{15}$ solar masses show that only
$\sim 6\%$ of them seem to have nonthermal activity with the same spectral
features observed in the Coma cluster. This number should be compared with 
the statistics of radio halos (Feretti et al. 2000) which seems to suggest 
that $\sim 30\%$ of the clusters with X-ray luminosity larger than $10^{45}$ 
erg/s have such radio halos. This comparison should however be taken with 
caution. In fact, if the radio halos found by Feretti et al. (2000) have 
steep spectra, then our statistics increases appreciably and the disagreement
may be attenuated. Unfortunately the spectrum is not available for all radio halos.
A more detailed analysis of recent mergers accounting in detail for the time
dependent electron losses is being currently carried out (Gabici \& Blasi, in 
preparation). 

We can summarize our conclusions as follows: 
\begin{itemize}
\item[1)] the diffuse nonthermal activity should not be directly associated to 
protons accelerated at merger shocks within the cluster volume;
\item[2)] the nonthermal activity should not correlate directly with major 
cluster mergers, unless the turbulence induced by mergers is responsible for 
particle acceleration; if a correlation is confirmed between radio halos and 
clusters that suffered major mergers, as seems to emerge from the analysis
of Buote et al. (2001), then the natural conclusion is that the nonthermal 
particles are not accelerated at shocks but rather energized by other processes,
possibly related to resonant wave-particle interactions;
\item[3)] if electrons directly accelerated at merger shocks are the sources of 
radio halos and HXR emission, then only about $6\%$ of the clusters with mass 
$10^{15}$ solar masses are expected to have such nonthermal activity (note that
even in these cases the acceleration is not expected to occur at shocks formed
during major mergers).
\end{itemize}

\acknowledgments 
We are grateful to G. Brunetti and A. Diaferio for many instructive 
discussions on the topics of nonthermal radiation in clusters of galaxies 
and velocity dispersions of clusters in overdense regions respectively.
We want also to thank the anonymous referee for his comments that helped
us to improve the present paper.

\end{document}